\documentclass{article}

\usepackage{latexsym}
\usepackage{epsfig}
\usepackage{amssymb}
\usepackage{times}
\usepackage[BCAY]{apacite}

\begin{document} 

\title{Self-organization of the Sound Inventories: Analysis and Synthesis of the Occurrence and Co-occurrence Networks 
of Consonants}

\author{Animesh Mukherjee$^1$, Monojit Choudhury$^2$,\\ Anupam Basu$^1$ and Niloy Ganguly$^1$\\$^1$Department of 
Computer Science and Engineering,\\ Indian Institute of Technology, Kharagpur, India -- 721302\\$^2$Microsoft Research 
India, Bangalore -- 560080}

\maketitle

\begin{abstract}
The sound inventories of the world's languages self-organize themselves giving rise to similar cross-linguistic 
patterns. In this work we attempt to capture this phenomenon of self-organization, which shapes the structure of the 
consonant inventories, through a complex network approach. For this purpose we define the occurrence and co-occurrence 
networks of consonants and systematically study some of their important topological properties. A crucial observation 
is that the occurrence as well as the co-occurrence of consonants across languages follow a power law distribution. 
This property is arguably a consequence of the principle of preferential attachment. In order to support this argument 
we propose a synthesis model which reproduces the degree distribution for the networks to a close approximation. We 
further observe that the co-occurrence network of consonants show a high degree of clustering and subsequently refine 
our synthesis model in order to incorporate this property. Finally, we discuss how preferential attachment manifests 
itself through the evolutionary nature of language. 
\end{abstract}


\section{Introduction}\label{intro}

Sound inventories of human languages show a considerable extent of symmetry. This symmetry is primarily a reflection 
of the self-organizing behavior that goes on in shaping the structure of the inventories~\cite{Oudeyer:06}. It has 
been postulated previously that such a self-organizing behavior can be explained through the principles of functional 
phonology, namely, {\em maximal perceptual contrast}~\cite{Lindblom:88}, {\em ease of 
articulation}~\cite{Boer:00,Lindblom:88}, and {\em ease of learnability}~\cite{Boer:00}. These explanations are an 
outcome of the {\em macroscopic} level~\cite{Arhem:04} of analysis, and are quite commonly used by traditional 
linguists~\cite{Boersma:98,Clements:04,Flemming:02,Lindblom:88,Trub:39}. Of late, it has been shown that even in the 
{\em microscopic} level~\cite{Arhem:04}, the emergent behavior of the vowel inventories in particular, can be 
satisfactorily explained through multi-agent simulation models~\cite{Boer:00}. However, instances of such modeling 
either in the micro or in the {\em mesoscopic} levels~\cite{Arhem:04} to demonstrate the organizational principles of 
the consonant inventories, are absent in literature.       

In this work we present a mesoscopic model, grounded on the theories of {\em complex networks} (for a review 
see~\cite{Albert:02,Newman:03}), in order to capture the self-organizing principles of the consonant inventories. We 
call our model mesoscopic since it does not make use of the functional properties of the macro level for explaining 
the structure of the consonant inventories; nor does it incorporate microscopic interactions between the speakers of a 
language (usually modeled through linguistic agents~\cite{Boer:00}), in order to reproduce this structure. We rather 
base our model on slightly coarse grained components like languages and consonants, and study the interactions between 
and within them respectively. In order to capture these interactions we define two networks namely, {\bf PlaNet} or 
the {\bf P}honeme {\bf La}nguage {\bf Net}work and {\bf PhoNet} or the {\bf Pho}neme Phoneme {\bf Net}work. PlaNet is 
a {\em bipartite} network which has two sets of nodes, one labeled by the languages while the other by the consonants. 
Edges run between the nodes of these two sets depending on whether or not a particular consonant occurs in a 
particular language. On the other hand, PhoNet is the one-mode projection\footnote{From a bipartite network, one can 
construct its unipartite
counterpart, the so-called one-mode projection onto actors, as a network consisting solely of the social actors
as nodes, two of which are connected by an edge for each social tie they both participate in. For example, two 
consonant nodes in the one-mode projection are connected as many times as they have co-occurred across the language 
inventories.} of PlaNet onto the consonant nodes. Hence PhoNet is a weighted {\em unipartite} network of consonants 
where an edge between two nodes signifies their co-occurrence likelihood over the consonant inventories. The 
construction of PlaNet, and subsequently PhoNet, are motivated by similar modeling of various complex phenomena 
observed in nature and society, such as,
\begin{itemize}
\item Movie-actor network, where movies and actors constitute the two partitions and an edge between them signifies 
that a particular actor acted in a particular movie \cite{Ramasco:04}. In the corresponding one-mode projection onto 
the actor nodes, any two actors are connected as many times as they have co-acted in a movie. 
\item Article-author network, where the edges denote which person has authored which articles~\cite{Newman:01}. In 
this case the one-mode projection onto the author nodes comprises of a pair of authors connected as many times as they 
have co-authored an article. 
\item Metabolic network of organisms, where the corresponding partitions are chemical compounds and metabolic 
reactions. Edges run between partitions depending on whether a particular compound is a substrate or result of a 
reaction~\cite{Jeong:00}. In this case the one-mode projection onto the chemical compounds comprises a pair of 
compounds connected by as many edges as they have co-participated in a metabolic reaction.
\end{itemize}
Modeling of the complex systems referred above as networks has proved to be a comprehensive and emerging way of 
capturing the underlying generating mechanisms of such systems~\cite{Albert:02,Newman:03}. In this direction there 
have been some attempts as well to model the intricacies of human languages through complex networks. Word networks 
based on synonymy~\cite{Yook:01}, co-occurrence \cite{Cancho:01}, and phonemic edit-distance~\cite{Vitevitch:05} are 
examples of such attempts. The present work also uses the concept of complex networks to develop a platform for a 
holistic analysis as well as synthesis of the distribution of the consonants across the languages.  

In the current work we present some of the exciting properties of the consonant inventories through the analysis as 
well as synthesis of PlaNet and PhoNet. A significant property we observe is that the consonant nodes in PlaNet as 
well as PhoNet have a power law {\em degree distribution} with an exponential cut-off. This property is arguably a 
consequence of the principle of {\em preferential attachment}~\cite{Albert:99}. In order to support this argument we 
present a synthesis model for PlaNet which generates the degree distribution of the consonant nodes and mimics the 
real data to a very close approximation. However, though the degree distributions of the empirical and the synthesized 
PlaNet match closely, their one-mode projections (the empirical and the synthesized PhoNet) seem to differ especially 
in their {\em clustering coefficients}~\cite{Watts:98}. The clustering coefficient of the empirical PhoNet is 
substantially higher than that of the synthesized version, indicating that consonants tend to frequently occur in 
cohesive groups or communities~\cite{Mukherjee:06}. We therefore modify the synthesis model and allow {\em triad} 
(i.e., fully connected triplets) formation. As a consequence of this modification the degree distributions of PlaNet 
and PhoNet as well as the clustering coefficient of PhoNet match their respective synthesized versions with a very 
high accuracy. 

The rest of the article is structured as follows. In section~\ref{def_con} we formally define PlaNet and PhoNet and 
outline their construction procedure. We also present some interesting studies pertaining to the structural properties 
of PlaNet as well as PhoNet in the same section. In section~\ref{synth} we propose a synthesis model for PlaNet based 
on the principle of preferential attachment. We also identify some necessary refinements in the synthesis model (in 
order to improve upon the average clustering coefficient of PhoNet) and subsequently extend it to incorporate these 
refinements in the same section. Finally we conclude in section~\ref{conc} by summarizing our contributions, pointing 
out some of the implications of the current work and indicating the possible future directions.  

\section{Definition, Construction and Analysis of PlaNet and PhoNet}\label{def_con}

In this section we formally define PlaNet and PhoNet followed by a description of their construction procedure. We 
also present some interesting studies pertaining to the topological properties of these networks.
 
\subsection{Definition}

\begin{figure}
\begin{center}
\includegraphics[width=3.2in]{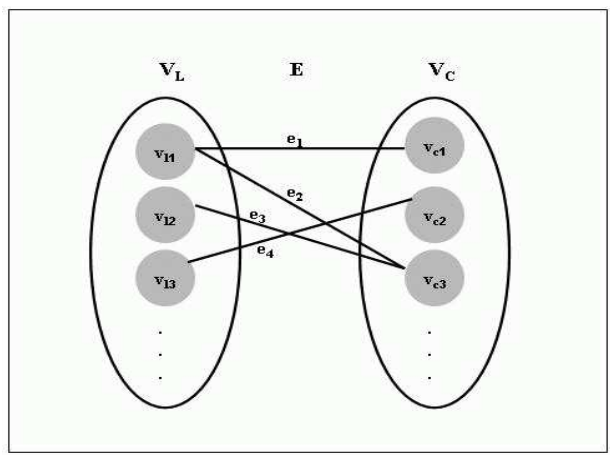}
\caption{Illustration of the nodes and edges of PlaNet.}
\label{planet}
\end{center}
\end{figure}

{\bf Definition of PlaNet:} We define PlaNet (the network of consonants and languages) as a bipartite graph G = 
$\langle${V$_L$, V$_C$, E}$\rangle$ where V$_L$ is the set of {\em nodes} labeled by the languages and V$_C$ is the 
set of nodes labeled by the consonants. E is the set of edges that run between V$_L$ and V$_C$. There is an {\em edge} 
$e$ $\in$ E between two nodes $v_l$ $\in$ V$_L$ and $v_c$ $\in$ V$_C$ if and only if the consonant {\em c} occurs in 
the language {\em l}. Figure~\ref{planet} illustrates the nodes and edges of PlaNet.

\begin{figure}
\begin{center}
\framebox{
\includegraphics[width=3.2in]{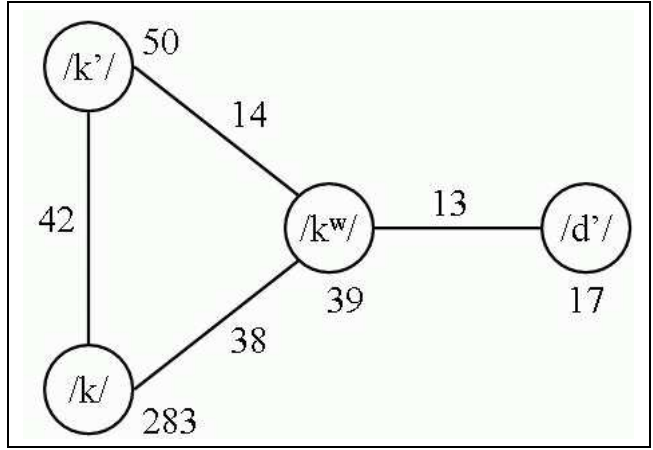}
}
\caption{A partial illustration of the nodes and edges in PhoNet. The labels of the nodes denote the consonants 
represented in IPA (International Phonetic Alphabet). The numerical values against the edges and nodes represent their 
corresponding weights. For example /$k$/ occurs in 283 languages; /$k^w$/ occurs in 39 languages while they co-occur 
in 38 languages.}
\label{graph}
\end{center}
\end{figure}

{\bf Definition of PhoNet:} PhoNet, which is the one-mode projection of PlaNet (projection taken on the consonant 
nodes), can be defined as a network of consonants represented by a graph G = $\langle$ V$_C$, E $\rangle$ where V$_C$ 
is the set of nodes labeled by the consonants and E is the set of all the edges in G. There is an edge $e$ $\in$ E 
between two nodes, if and only if there exists one or more language(s) where the nodes (read consonants) co-occur. The 
weight of the edge $e$ (also {\em edge-weight}) is the number of languages in which the consonants connected by $e$ 
co-occur. Figure~\ref{graph} presents a partial illustration of PhoNet.

\subsection{Construction} 

{\bf Construction of PlaNet:} Many typological studies~\cite{Hinskens:03,Ladefoged:96,Lindblom:88} of segmental 
inventories have been carried out in past on the UCLA Phonological Segment Inventory Database 
(UPSID)~\cite{Maddieson:84}. UPSID records the sound inventories of 317 languages covering all the major language 
families of the world. In this work we have used UPSID comprising of these 317 languages and 541 consonants found 
across them, for constructing PlaNet. Consequently, there are 317 elements (nodes) in the set V$_L$ and 541 elements 
(nodes) in the set V$_C$. The number of elements (edges) in the set E as computed from PlaNet is 7022. At this point 
it is important to mention that in order to avoid any confusion in the construction of PlaNet we have appropriately 
filtered out the {\em anomalous} and the {\em ambiguous} segments~\cite{Maddieson:84} from it. In UPSID, a segment has 
been classified as anomalous if its existence is doubtful and ambiguous if there is insufficient information about the 
segment. For example, the presence of both the palatalized dental plosive and the palatalized alveolar plosive are 
represented in UPSID as palatalized dental-alveolar plosive. According to popular techniques~\cite{Peri:02}, we have 
completely ignored the anomalous segments from the data set, and included the ambiguous ones as separate segments 
because there are no descriptive sources explaining how such ambiguities might be resolved.  

{\bf Construction of PhoNet:} Once PlaNet is constructed as described above, we can easily construct PhoNet by taking 
an one-mode projection of PlaNet onto the consonant nodes. Consequently, the set V$_C$ for PhoNet comprises of 541 
elements (nodes) and the set E comprises of 34012 elements (edges). 

\subsection{Degree Distribution}

The {\em degree} of a node $u$, denoted by $k_u$ is defined as the number of edges connected to $u$. The term {\em 
degree distribution} is used to denote the way degrees ($k_u$) are distributed over the nodes ($u$).
 
\subsubsection{Degree Distribution of PlaNet}

Since PlaNet is bipartite in nature it has two degree distribution curves one corresponding to the nodes in the set 
V$_L$  and the other corresponding to the nodes in the set V$_C$.

\begin{figure}
\begin{center}
\includegraphics[width=3.2in]{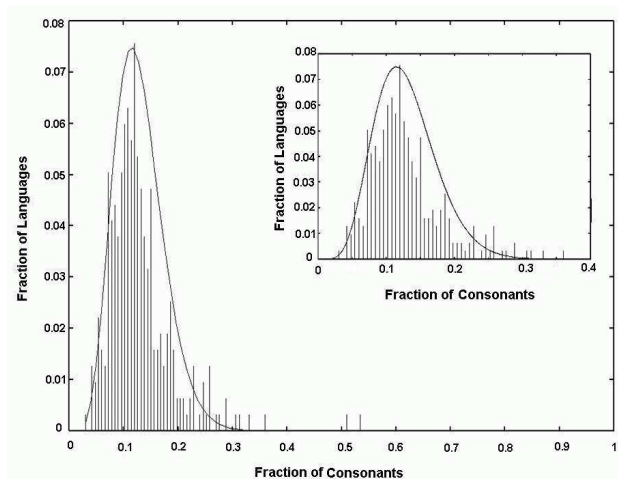}
\caption{Degree distribution of PlaNet for the set V$_L$. The figure in the inner box is a magnified version of a 
portion of the original figure.}
\label{dd_lang}
\end{center}
\end{figure}

{\bf {Degree distribution of the nodes in V$_L$:}} Figure~\ref{dd_lang} shows the degree distribution of the nodes in 
V$_L$ where the x-axis denotes the degree of each node expressed as a fraction of the maximum degree and the y-axis 
denotes the number of nodes having a given degree  expressed as a fraction of the total number of nodes in V$_L$. 

Figure~\ref{dd_lang} indicates that the number of consonants appearing in different languages follow a 
$\beta$-distribution\footnote{A random variable is said to have a $\beta$-distribution with parameters $\alpha$ $>$ 0 
and $\beta$ $>$ 0 if and only if its probability mass function is given by,\\
$f(x) = \frac{\Gamma({\alpha} + {\beta})}{{\Gamma} ({\alpha}) {\Gamma} ({\beta})}x^{\alpha-1}(1-x)^{\beta-1}$\\ 
for 0 $<$ x $<$ 1 and $f(x)$ = 0 otherwise. $\Gamma$($\cdot$) is the Euler's gamma function.} (see~\cite{Bulmer:79} 
for reference) which is right skewed with the values of $\alpha$ and $\beta$ equal to 7.06 and 47.64 (obtained using 
maximum likelihood estimation method) respectively. This asymmetry in the distribution points to the fact that 
languages usually tend to have smaller consonant inventory size, the best value being somewhere between 10 and 30. The 
distribution peaks roughly at 21 (which is its mode) whereas the mean of the distribution is 22 indicating that on an 
average the languages in UPSID have a consonant inventory of 22~\cite{Maddieson:99}. 

\begin{figure}[!t]
\begin{center}
\includegraphics[width=3.3in]{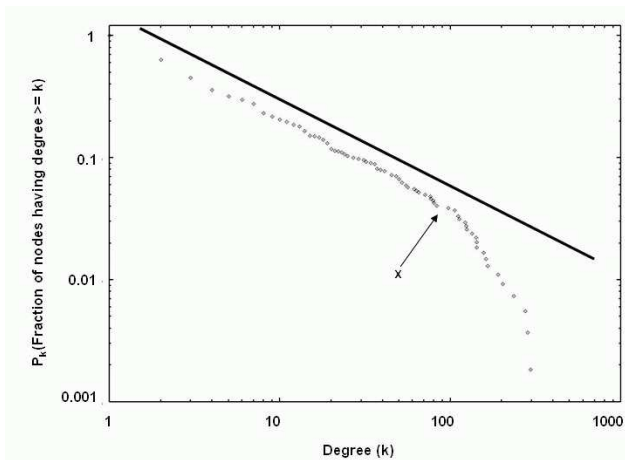}
\caption{Degree distribution of PlaNet for the set V$_C$ in a log-log scale. The letter {\bf x} denotes the cut-off 
point.}
\label{dd_cons}
\end{center}
\end{figure}

{\bf {Degree distribution of the nodes in V$_C$:}} Figure~\ref{dd_cons} illustrates the degree distribution plot for 
the nodes in V$_C$ in log-log scale. In this figure the x-axis represents the degree ($k$)  and the y-axis represents 
$P_k$, where $P_k$ is the fraction of nodes having degree greater than or equal to $k$. 

Figure~\ref{dd_cons} clearly shows that the curve follows a power law distribution with an exponential cut-off. The 
cut-off point is indicated by the letter {\bf x} in the figure. We find that there are 22 consonant nodes which have 
their degree above the cut-off range. However, the remaining consonant nodes of PlaNet exhibit a power law degree 
distribution of the form 
\begin{equation}\label{plaw}
y = Ax^{-\gamma}
\end{equation}
The values of the parameters A and $\gamma$ in both the figures, as computed by the least square error method, are 
noted in Table~\ref{param}.

\begin{table*}\centering
\caption{The values of the parameters A and $\gamma$.}
\begin{tabular}{|l|l|l|}
\cline{1-3}
\vbox to1.88ex{\vspace{1pt}\vfil\hbox to12.80ex{\hfil Parameter\hfil}} & 
\vbox to1.88ex{\vspace{1pt}\vfil\hbox to24.60ex{\hfil  Figure~\ref{dd_cons}\hfil}} & 
\vbox to1.88ex{\vspace{1pt}\vfil\hbox to24.00ex{\hfil Figure~\ref{phonet_deg}\hfil}} \\

\cline{1-3}
\vbox to1.88ex{\vspace{1pt}\vfil\hbox to12.80ex{\hfil A\hfil}} & 
\vbox to1.88ex{\vspace{1pt}\vfil\hbox to24.60ex{\hfil 1.04\hfil}} & 
\vbox to1.88ex{\vspace{1pt}\vfil\hbox to24.00ex{\hfil 41.26\hfil}} \\

\cline{1-3}
\vbox to1.88ex{\vspace{1pt}\vfil\hbox to12.80ex{\hfil $\gamma$\hfil}} & 
\vbox to1.88ex{\vspace{1pt}\vfil\hbox to24.60ex{\hfil 0.71\hfil}} & 
\vbox to1.88ex{\vspace{1pt}\vfil\hbox to24.00ex{\hfil 0.89\hfil}} \\

\cline{1-3}
\end{tabular}
\label{param}
\end{table*}

\subsubsection{Degree Distribution of PhoNet}

\begin{figure}
\begin{center}
\includegraphics[width=3.3in]{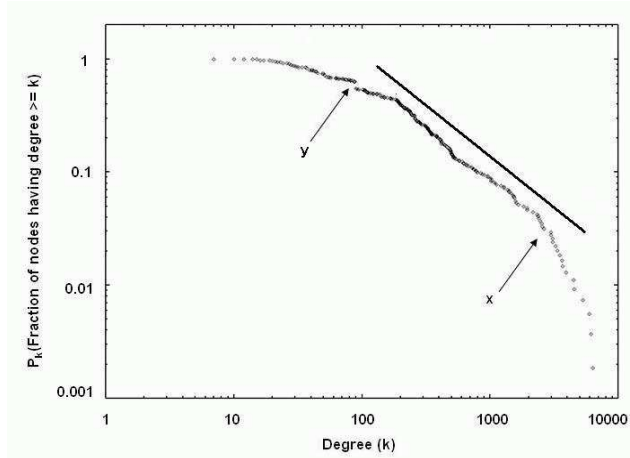}
\caption{Degree distribution of the nodes in PhoNet in a log-log scale. The letters {\bf x} and {\bf y} denote the 
cut-off points.}
\label{phonet_deg}
\end{center}
\end{figure}

Since PhoNet is a weighted network, we report the distribution of the {\em weighted degree} of its nodes. The weighted 
degree $k$ for a node $i$ can be defined as~\cite{Barrat:04},
\begin{equation}
k = \sum_{\forall j}w_{ij}
\end{equation}
where $j$ is a neighbor of $i$ in the network and $w_{ij}$ is the weight of the edge connecting the nodes $i$ and $j$. 
Figure~\ref{phonet_deg} shows the degree distribution curve for PhoNet in log-log scale. In this figure the x-axis 
represents the weighted degree ($k$) and the y-axis represents $P_k$, where $P_k$ is the fraction of nodes having 
degree greater than or equal to $k$. Interestingly, the degree distribution of the nodes in PhoNet show two different 
cut-off points marked by the letters {\bf x} and {\bf y} in the figure along with a power law region spanning from 
{\bf x} to {\bf y}. Through inspection we find that there are 15 consonants above the cut-off {\bf x} which indicates 
that these consonants co-occur very frequently exhibiting an hub-like nature~\cite{Mukherjee:06,Newman:03}. On the 
other hand, the  degree distribution of the nodes up to the cut-off point {\bf y} is approximately a straight line 
indicating that the rate of change of $P_k$ with respect to $k$ is quite slow in this region. The reason behind this 
behavior is that there are only a negligibly small fraction of nodes that exist in this region which, causes almost no 
increase in the cumulative fraction $P_k$. However, the rest of the consonant nodes (spanning from the point {\bf y} 
to the point {\bf x}) show a power law degree distribution of the form specified in equation~\ref{plaw}. 
Table~\ref{param} reports the values of the parameters A and $\gamma$.

In most of the networked systems like the society, the Internet, the World Wide Web, and many others, power law degree 
distribution emerges for the phenomenon of preferential attachment, i.e., when ``the rich get richer"~\cite{Simon:55}. 
With reference to PlaNet and PhoNet this preferential attachment can be interpreted as the tendency of a language to 
choose a consonant that has been already chosen by a large number of other languages. In order to validate the above 
argument, in section~\ref{synth}, we present a synthesis model based on preferential attachment (where the 
distribution of the consonant inventory size is known {\em a priori}) that mimics the occurrence and co-occurrence 
distributions of the consonant nodes to a very close approximation. 

\subsection{Clustering Coefficient}

The weighted clustering coefficient (in the one-mode projection PhoNet) for a node $i$ is defined as~\cite{Barrat:04},
\begin{equation}\label{cc} c_i = \frac{1}{\left(\sum_{\forall j}w_{ij}\right)(k_i-1)}{\sum_{\forall j, 
l}\frac{(w_{ij}+w_{il})}{2}a_{ij}a_{il}a_{jl}}\end{equation} where $j$ and $l$ are neighbors of $i$; $k_i$ represents 
the plain degree of the node $i$; $w_{ij}$, $w_{jl}$ and $w_{il}$ denote the weights of the edges connecting nodes $i$ 
and $j$, $j$ and $l$, and $i$ and $l$ respectively; $a_{ij}$, $a_{il}$, $a_{jl}$ are boolean variables indicating 
whether or not there is an edge between the nodes $i$ and $j$, $i$ and $l$, and $j$ and $l$ respectively. This 
coefficient is a measure of the local cohesiveness that takes into account the importance of the clustered structure 
on the basis of the amount of traffic or interaction intensity actually found on the local triplets. The parameter 
$c_i$ in equation~\ref{cc} counts for each triplet formed in the neighborhood of the vertex $i$, the weight of the two 
participating edges of the vertex $i$. The normalization factor $\left(\sum_{\forall j}w_{ij}\right)(k_i-1)$ accounts 
for the weight of each edge times the maximum possible number of triplets in which it may participate, and it ensures 
that $0~\le~c_i~\le~1$. Consequently, the average weighted clustering coefficient is given by, \begin{equation}c_{av} 
= \frac{1}{N}{\sum_{i=1}^{N}c_i}\end{equation} where, $N$ is the number of nodes in PhoNet. 

\begin{figure}
\begin{center}
\includegraphics[width=3.3in]{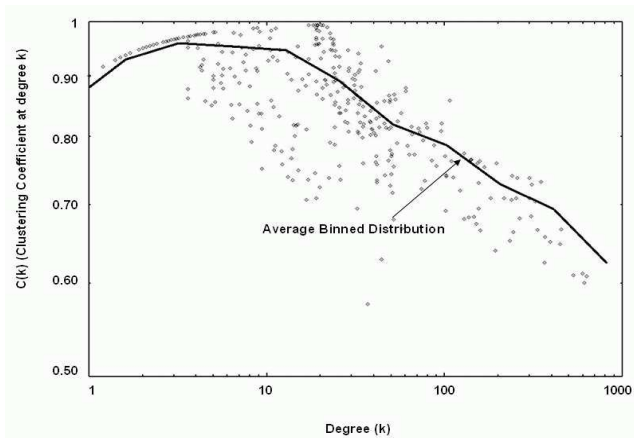}
\caption{Degree versus clustering coefficient for the nodes of PhoNet in log-log scale. The bold line shows the 
average binned distribution with bin sizes expressed in powers of 2.}
\label{dd_cc}
\end{center}
\end{figure}

The value of $c_{av}$ for PhoNet is 0.89 which indicates a huge clustering among the nodes of the network. This is 
primarily due to the fact that in PhoNet the consonant nodes tend to occur highly in cohesive groups or communities as 
shown by Mukherjee {\em et al.}~\cite{Mukherjee:06}. In order to further investigate how the clustering coefficient is 
related to the weighted degree of the nodes of PhoNet we plot the degree-dependent clusterings in Figure~\ref{dd_cc} 
in log-log scale. The figure shows a scatter plot as well as an average binned distribution with bin sizes expressed 
as powers of 2. It is quite interesting to observe that like many other social networks~\cite{Alava:06} in this case 
also the clustering is substantial (very close to one) for vertices with small degrees and gets lower with an 
increasing $k$. This is because the more neighbors a consonant node has the less probable it will be for those 
neighbors to co-occur in the same language and hence be connected with each other. 

The power law degree distributions observed for both PlaNet and PhoNet are indicative of the presence of preferential 
attachment in the organization of the consonant inventories. We are however interested in estimating the amount of 
preferential attachment involved to get an accurate view of the underlying dynamics guiding the formation of the 
consonant inventories. This can be best understood through a synthesis model which, we present in the next section 
that mimics the empirical data to a high precision.  
  	
\section{Synthesis Model Based on Preferential Attachment}\label{synth}

\begin{figure}[!t]
\begin{center}
\framebox{
\includegraphics[width=3.3in]{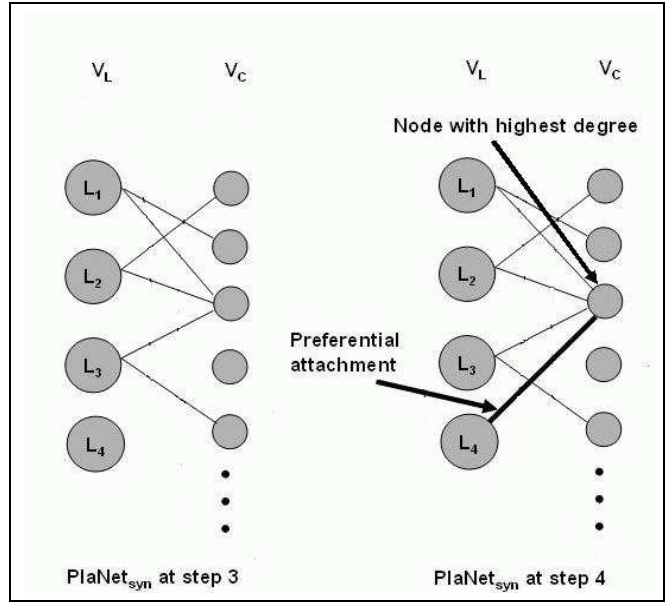}}
\caption{A partial step of the synthesis process. When the language L$_4$ has to connect itself with one of the nodes 
in the set V$_C$ it does so with the one having the highest degree (=3) rather than with others in order to achieve 
preferential attachment which is the working principle of our algorithm.}
\label{planetsyn}
\end{center}
\end{figure}

In this section we present a synthesis model for the PlaNet (henceforth PlaNet$_{syn}$) based on preferential 
attachment where the distribution of the consonant inventory size is assumed to be known {\em a priori}. Let V$_L$ = 
\{L$_1$,L$_2$,...,L$_{317}$\} have cardinalities (consonant inventory size) \{k$_1$,k$_2$,...,k$_{317}$\} 
respectively. We assume that the consonant nodes (V$_C$) of PlaNet$_{syn}$ are {\em unlabeled} (i.e., they are not 
labeled by a set of articulatory/acoustic features~\cite{Mukherjee:06} that characterizes them). We next sort the 
nodes L$_1$ through L$_{317}$ in ascending order of their cardinalities. At each time step a node L$_j$, chosen in 
order, preferentially attaches itself with k$_j$ {\em distinct} nodes (call each such node C$_i$) of the set V$_C$. 
The probability $Pr(C_i)$ with which the node L$_j$ attaches itself to the node C$_i$ is given by, 
\begin{equation}\label{pref}
Pr(C_i) = \frac{{d_{i}}^{\alpha} + \epsilon}{\sum_{{i^{'}} \in V^{'}_{C}} ({d_{i^{'}}}^{\alpha} + \epsilon)}
\end{equation} where, $d_i$ is the current degree of the node $C_i$, V$^{'}_{C}$ is the set of nodes in V$_C$ that are 
not already connected to L$_j$ and $\epsilon$ is the smoothing parameter which facilitates attachments to consonant 
nodes that have a degree close or equal to zero. The above process is repeated until all the language nodes L$_j \in 
$V$_L$ get connected to k$_j$ consonant nodes. Algorithm~1 summarizes the synthesis process and Figure~\ref{planetsyn} 
illustrates a partial step of this process. 

\begin{table}[h]\centering
\begin{tabular}{l}
\hline
{\bf Algorithm 1: The synthesis process}\\
\hline
\hline
{\bf Input:} Nodes L$_1$ through L$_{317}$ sorted in ascending\\
\hspace*{33pt}order of their cardinalities;\\ 
{\bf for} $t$ = 1 to 317 \{\\
\hspace*{10pt}Choose (in order) a node L$_j$ with cardinality k$_j$;\\
\hspace*{10pt}{\bf for} $c$ = 1 to k$_j$ \{\\
\hspace*{20pt}Connect L$_j$ to a node C$_i$ $\in$ V$_C$ to which it is\\
\hspace*{20pt}not yet connected, following the distribution,\\
\hspace*{20pt} $Pr(C_i) = \frac{{d_{i}}^{\alpha} + \epsilon}{\sum_{{i^{'}} \in V^{'}_{C}} ({d_{i^{'}}}^{\alpha} + 
\epsilon)}$\\ 
\hspace*{20pt}where V$^{'}_{C}$ is the set of nodes in V$_C$ to which\\
\hspace*{20pt}L$_j$ is not yet connected;\\ 
\hspace*{10pt}\}\\  
\}\\
\hline
\end{tabular}
\end{table} 

As we shall see shortly, the aforementioned preferential attachment based model is able to explain the distribution of 
the consonants over languages reasonably well. However, at this point it would be worthwhile to mention the reason 
behind sorting the language nodes in ascending order of their cardinalities. With each consonant is associated two 
different frequencies; a)~the frequency of occurrence of a consonant over languages or the {\em type} frequency, and 
b)~the frequency of usage of the consonant in a particular language or the {\em token} frequency. Researchers have 
shown in the past that these two frequencies are positively correlated. Nevertheless, our synthesis model based on 
preferential attachment takes into account only the type frequency of a consonant and not its token frequency. If 
language is considered to be an evolving system~\cite{Lightfoot:99} then both of these frequencies, in one generation, 
should play an important role in shaping the inventory structure of the next generation. 

In the later stages of our synthesis process when the attachments are strongly preferential, the type frequencies span 
over a large range and automatically guarantee the token frequency (since they are positively correlated). However, in 
the initial stages of this process the attachments that take place are random in nature and therefore the type 
frequencies of all the nodes are roughly equal. At this point it is the token frequency (absent in our model) that 
should discriminate between the nodes. This error due to the loss of information about the token frequency in the 
initial steps of the synthesis process can be minimized by allowing only a small number of attachments (so that there 
is less spreading of the error). This is primarily the reason why we sort the language nodes in the ascending order of 
their cardinalities so that there are only a few random connections in the initial steps resulting in minimum error 
propagation. 

Apart from the ascending order, we have also simulated the model with descending and random order of the inventory 
size. The degree distribution obtained by considering ascending order of the inventory size, matches more accurately 
than in the other two scenarios.
    
\begin{figure}
\begin{center}
\includegraphics[width=3.3in]{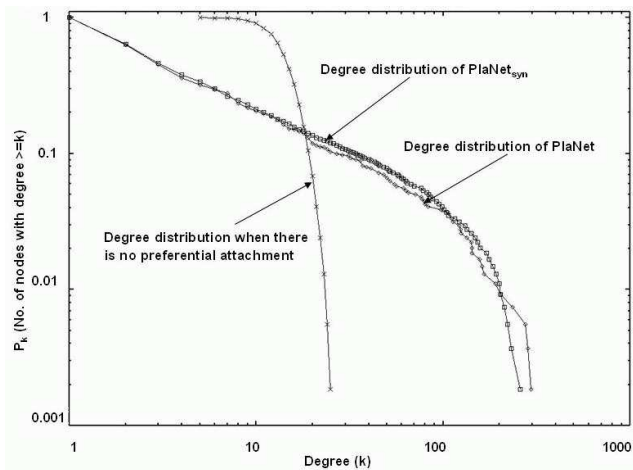}
\caption{Degree distribution of the nodes in V$_C$ for PlaNet$_{syn}$, PlaNet, as well as when the model incorporates 
no preferential attachment; for PlaNet$_{syn}$, $\epsilon$~=~0.5, $\alpha$~=~1.44, and the results are averaged over 
100 simulation runs.}
\label{deg_synth}
\end{center}
\end{figure}

{\bf Simulation Results:} Simulations reveal that in case of PlaNet$_{syn}$ the degree distribution of the nodes 
belonging to V$_C$ fit well with the empirical results we obtained earlier in section~\ref{def_con}. Good fits emerge 
for 0.4~$\leq~\epsilon~\leq$~0.6 and 1.4~$\leq~\alpha~\leq$~1.5 with the best being at $\epsilon$~=~0.5 and 
$\alpha$~=~1.44. Figure~\ref{deg_synth} shows the degree $k$ versus $P_k$ plots averaged over 100 simulation runs.

The mean error\footnote{Mean error is defined as the average difference between the ordinate pairs (say $y$ and 
$y^{'}$) where the abscissas are equal. In other words if there are $N$ such ordinate pairs then the mean error can be 
expressed as $\frac{\sum\mid y-y^{'}\mid}{N}$} between the degree distribution plots of PlaNet and PlaNet$_{syn}$ is 
$\sim$~0.01. The match degrades as we reduce the value of $\alpha$ to 1 (linear preferential attachment) where the 
mean error is 0.03. The match further degrades as we continue to reduce the value of $\alpha$ and is worst for 
$\alpha$~=~0 (i.e., there is no preferential attachment incorporated in the model and all connections are 
equiprobable) where the mean error is as high as 0.35. 

\begin{figure}
\begin{center}
\includegraphics[width=3.3in]{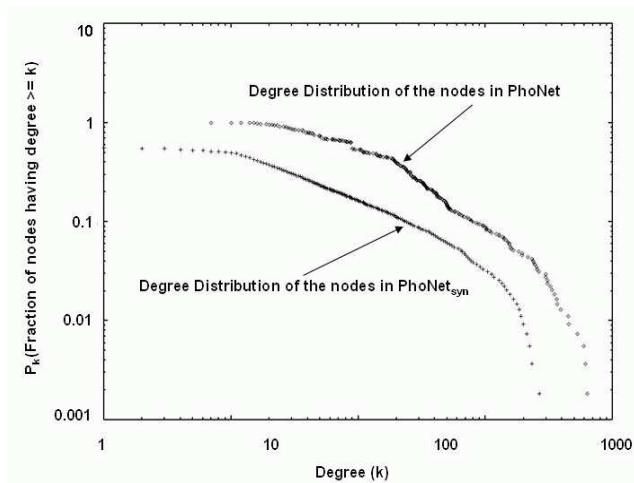}
\caption{Degree distribution of the nodes of PhoNet$_{syn}$ and PhoNet in log-log scale.}
\label{phd_syn}
\end{center}
\end{figure}

After having studied the properties of PlaNet$_{syn}$ as discussed above, we now construct PhoNet$_{syn}$ from it in 
the approach outlined in section~\ref{def_con}. Next, the degree distribution as well as the clustering coefficient of 
PhoNet$_{syn}$ is compared with that of PhoNet. Figure~\ref{phd_syn} illustrates the degree $k$ versus $P_k$ plot for 
PhoNet$_{syn}$ (along with that of PhoNet). It is clear from the figure that the degree distribution curves for PhoNet 
and PhoNet$_{syn}$ are qualitatively similar, although there is a significant amount of quantitative difference 
between the two (mean error $\sim$~0.45). Moreover, the average clustering coefficient $c^{'}_{av}$ of 
PhoNet$_{syn}$~(0.55) differs largely from that of PhoNet~(0.89).

Due to this large deviation from the real data, our synthesis model needs to be refined so that it not only mimics the 
degree distribution of PlaNet but also reproduces the degree distribution as well as the average clustering 
coefficient of PhoNet. The primary reason for this deviation in the results is that the high degree of clustering, 
observed in PhoNet, is not taken into account by our synthesis model. Such a high clustering is a consequence of the 
fact that apart from preferential attachment there is some other force governing the structure of the consonant 
inventories. Mukherjee {\em et al.}~\cite{Mukherjee:06} reports that this force tends to bind a set of consonants in 
cohesive groups which essentially leads to the emergence of a pattern of co-occurrence (resulting in high clustering) 
among them.  

Therefore, in order to boost up the clustering coefficient we refine the model to allow the formation of triads (i.e., 
fully connected triplets). This technique has been used by several models in the past~\cite{Ramasco:04}, which has led 
to increased clustering, closer to what is found in real networks. The refinement can be achieved in our model by 
having a language node L$_j$ attach to some node C$_i \in$ V$_C$ if it has already attached itself to a {\em neighbor} 
of C$_i$. Two consonant nodes C$_1$ and C$_2$ are neighbors if a language node (other than L$_j$) attaches itself to 
both C$_1$ and C$_2$ in an earlier step of the synthesis process. This phenomenon leads to the formation of a large 
number of triangles and/or triads in the one-mode projection, which in turn is expected to yield a higher clustering 
coefficient.   

In this model we denote the probability of triad formation by $p_t$. At each time step a language node L$_j$ (chosen 
from the set of language nodes sorted in ascending order of their degrees) makes the first connection to some 
consonant node C$_i$ $\in$ V$_C$ (L$_j$ is not already connected to C$_i$) preferentially following the distribution 
$Pr(C_i)$ (specified in equation~\ref{pref}). For the rest of the ($k_j$-1) connections the language node L$_j$ 
attaches itself preferentially to only the neighbors of C$_i$ (to which L$_j$ is not already connected) with a 
probability $p_t$. Consequently, L$_j$ connects itself preferentially to the non-neighbors of C$_i$ (to which L$_j$ is 
not already connected) with a probability (1-$p_t$). Accordingly, the neighbor set of $C_i$ gets updated.

\begin{figure*}[!t]
\begin{center}
\framebox{
\includegraphics[width=5in]{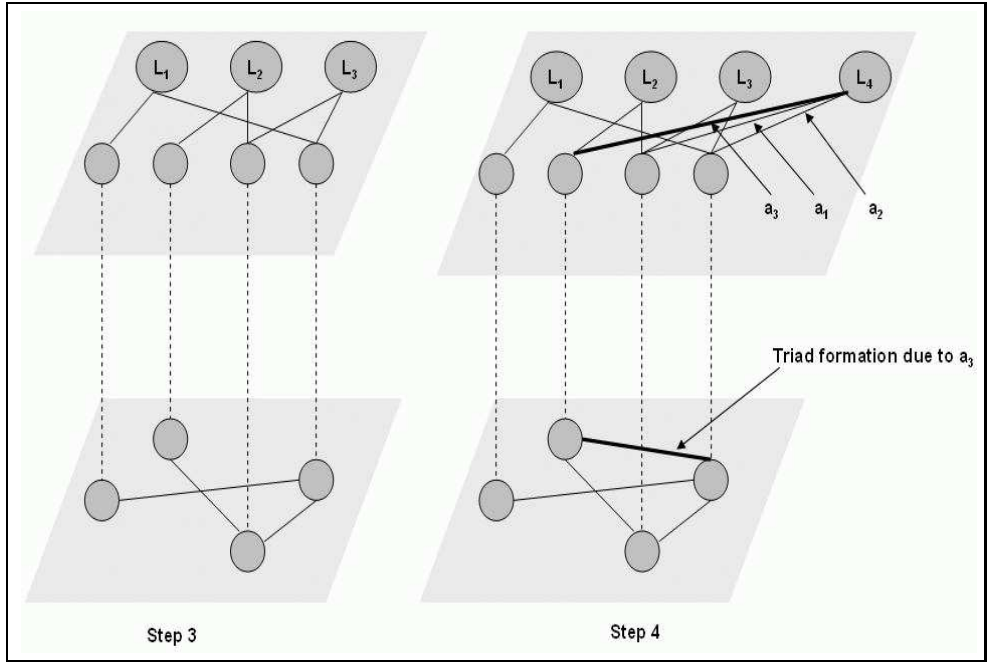}}
\caption{A partial step of the synthesis process. If the language node L$_4$ (which has degree 3) has the initial 
connection $a_1$ (due to preferential attachment) then according to the synthesis model the following connections 
would be $a_2$ and $a_3$ respectively in that order. This series of connections increases the co-occurrence by 
allowing formation of triads in the one-mode projection. The bold line in the figure indicates the edge that completes 
the triad.}
\label{triadsyn}
\end{center}
\end{figure*}

The entire idea mentioned above is summed up in Algorithm~2. Figure~\ref{triadsyn} shows a partial step of the 
synthesis process illustrated in Algorithm~2.

\begin{table}[!t]\centering
\begin{tabular}{l}
\hline
{\bf Algorithm 2: The refined synthesis process}\\
\hline
\hline
{\bf Input:} Nodes L$_1$ through L$_{317}$ sorted in ascending\\
\hspace*{33pt}order of their cardinalities;\\ 
{\bf for} $t$ = 1 to 317 \{\\
\hspace*{10pt}Choose (in order) a node L$_j$  with cardinality k$_j$;\\
\hspace*{10pt}Connect L$_j$ to a node C$_i$ $\in$ V$_C$ with which it is\\
\hspace*{10pt}not yet connected, following the distribution,\\ 
\hspace*{10pt} $Pr(C_i) = \frac{{d_{i}}^{\alpha} + \epsilon}{\sum_{{i^{'}} \in V^{'}_{C}} ({d_{i^{'}}}^{\alpha} + 
\epsilon)}$\\ 
\hspace*{10pt}where V$^{'}_{C}$ is the set of nodes in V$_C$ to which L$_j$\\
\hspace*{10pt}is not already connected;\\ 
\hspace*{10pt}{\bf for} $c$ = 2 to $k_j$ \{\\
\hspace*{20pt}Connect L$_j$ with a probability $p_t$ to a neigh-\\
\hspace*{20pt}bor C$^{'}_i$ of the node C$_i$ (to which L$_j$ is not yet\\
\hspace*{20pt}connected) following the distribution $Pr(C^{'}_i)$;\\
\hspace*{20pt}and,\\
\hspace*{20pt}Connect L$_j$ with probability (1-$p_t$) to a non-\\
\hspace*{20pt}neighbor C$^{''}_i$ of the node C$_i$ (to which L$_j$ is\\
\hspace*{20pt} not yet connected) following the distribution\\
\hspace*{20pt}$Pr(C^{''}_i)$ and accordingly expand the neighbor\\
\hspace*{20pt}list of C$_i$;\\
\hspace*{10pt}\}\\
\}\\
\hline
\end{tabular}
\end{table} 

\subsection{Estimation of Model Parameters}

The refined model discussed in Algorithm~2 involves three different parameters namely $p_t$, $\alpha$, and $\epsilon$. 
Thus at this point it becomes necessary to figure out how these model parameters interact and thereby influence the 
results. For this purpose we perform simulations of the model for different values of $p_t$ (in the range [0.7,0.95]), 
$\alpha$ (in the range [1.1,1.5]), and $\epsilon$ (in the range [0.2,0.4]) and study the pairwise relationships 
between errors that show up in (1)~the degree distribution of PlaNet$_{syn}$, (2)~the degree distribution of 
PhoNet$_{syn}$, and (3)~the average clustering coefficient of PhoNet$_{syn}$.  

\begin{figure}[!t]
\begin{center}
\includegraphics[width=3.3in]{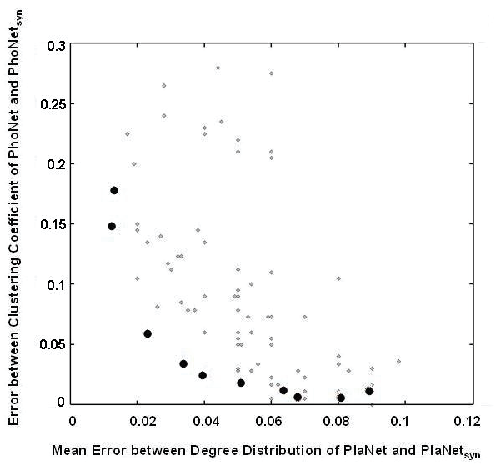}
\caption{The correlation of the mean error between the degree distribution of PlaNet/PlaNet$_{syn}$ with that of 
PhoNet/PhoNet$_{syn}$. Each point on the plot indicates a certain combination of the values of $p_t$, $\alpha$, and 
$\epsilon$. Some of the non-dominating points are indicated by the black circles.}
\label{pareto1}
\end{center}
\end{figure}

Figure~\ref{pareto1} shows that the mean error between the degree distribution of the pair PlaNet/PlaNet$_{syn}$ is 
negatively correlated to that of PhoNet/PhoNet$_{syn}$. Each point on the plot indicates a certain combination of the 
values of $p_t$, $\alpha$, and $\epsilon$. Some of the non-dominating~\cite{Ajith:05} points are indicated by black 
circles in the figure.

\begin{figure}[!t]
\begin{center}
\includegraphics[width=3.3in]{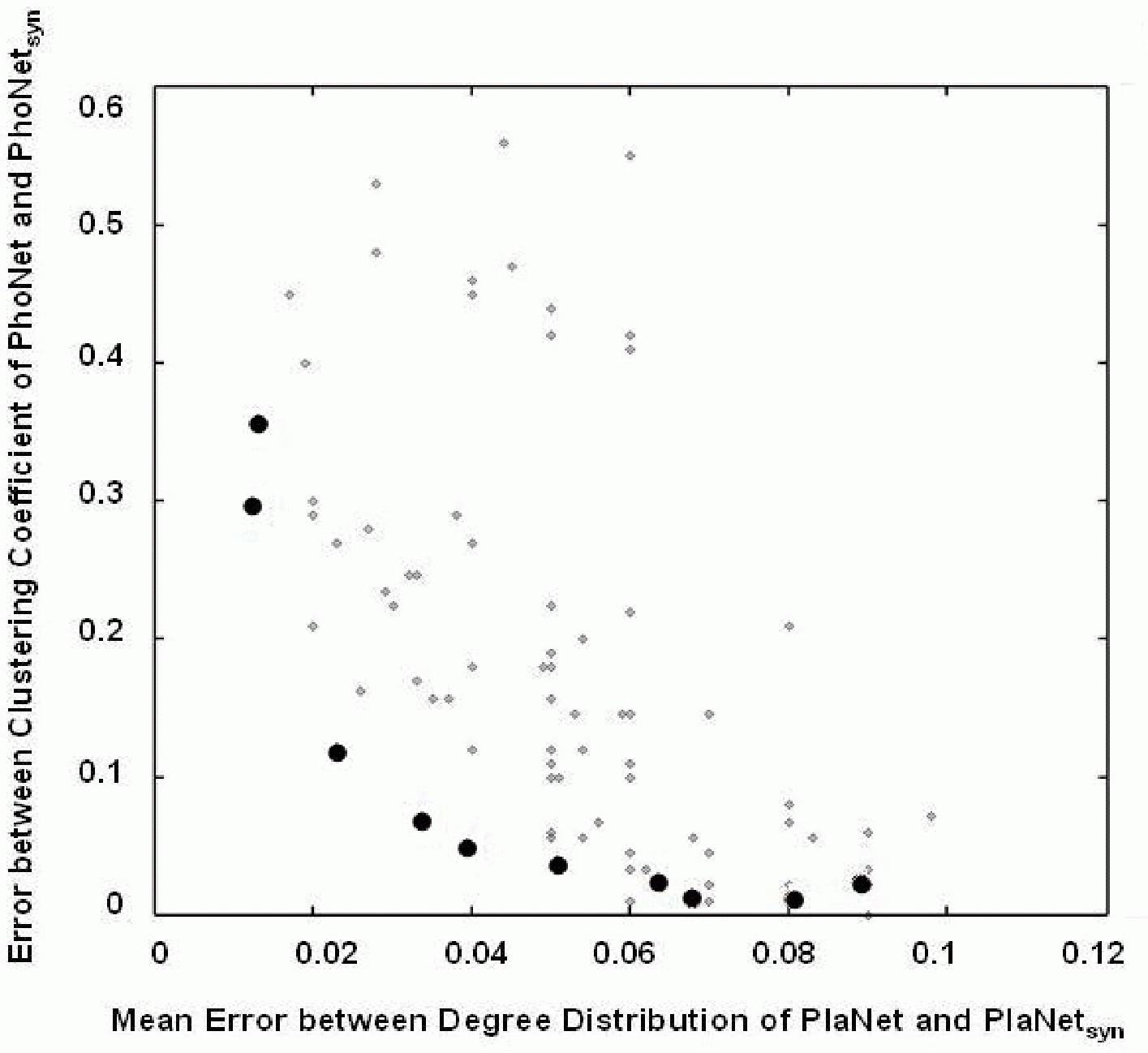}
\caption{The correlation of the mean error between the degree distribution of PlaNet/PlaNet$_{syn}$ with that of 
PhoNet/PhoNet$_{syn}$. Each point on the plot indicates a certain combination of the values of $p_t$, $\alpha$, and 
$\epsilon$. A few of the non-dominating points are marked by the black circles.}
\label{pareto2}
\end{center}
\end{figure}

Moreover, the mean error between the degree distribution of PlaNet/PlaNet$_{syn}$ is also negatively correlated to the 
error\footnote{The error in the average clustering coefficient is expressed as $\frac{\mid c_{av} - 
c^{'}_{av}\mid}{c_{av}}$ where $c_{av}$ and $c^{'}_{av}$ are the average clustering coefficients of PhoNet and 
PhoNet$_{syn}$ respectively.} between the average clustering coefficient of PhoNet/PhoNet$_{syn}$. 
Figure~\ref{pareto2} illustrates this negative correlation. In this case again each point on the plot indicates a 
certain combination of the values of $p_t$, $\alpha$, and $\epsilon$. Some of the non-dominating points are marked by 
black circles in the figure.

Further, the mean error between the degree distribution of PhoNet/PhoNet$_{syn}$ and the error between the average 
clustering coefficient of PhoNet/PhoNet$_{syn}$ do not show a direct dependence on one another (so that their effects 
could be assumed to be proportional), even though they are not strictly negatively correlated. 

\subsubsection{Combining the Objective Functions}

In a multi-objective scenario as discussed above, the most common technique, reported in literature~\cite{Ajith:05}, 
to arrive at a solution is to look for trade-offs rather than a single point that optimizes all the objective 
functions. One of the popular ways for such a trade-off is to combine (usually linearly) the objective functions 
together to result in a single objective function. In similar lines, we define the error $E$ which is the average of 
the mean error (say $M_E$) between degree distribution of PlaNet and PlaNet$_{syn}$, the mean error (say $M^{'}_E$) 
between the degree distribution of PhoNet and PhoNet$_{syn}$, and the error (say $C_E$) in the average clustering 
coefficient of PhoNet and PhoNet$_{syn}$. In other words,
\begin{equation}\label{erfc}
E = \frac{1}{3}\left(M_E + M^{'}_E + C_E\right)  
\end{equation}
It is to be noted that there can be multiple other ways of combining the above three errors and equation~\ref{erfc} is 
just one of them. Having defined the error $E$, we perform a detailed study of the parameter space in order to compute 
the minimum value of $E$.

\subsubsection{Experiment to Compute the Minimum Error}
  
\begin{figure}[!t]
\begin{center}
\includegraphics[width=3.4in]{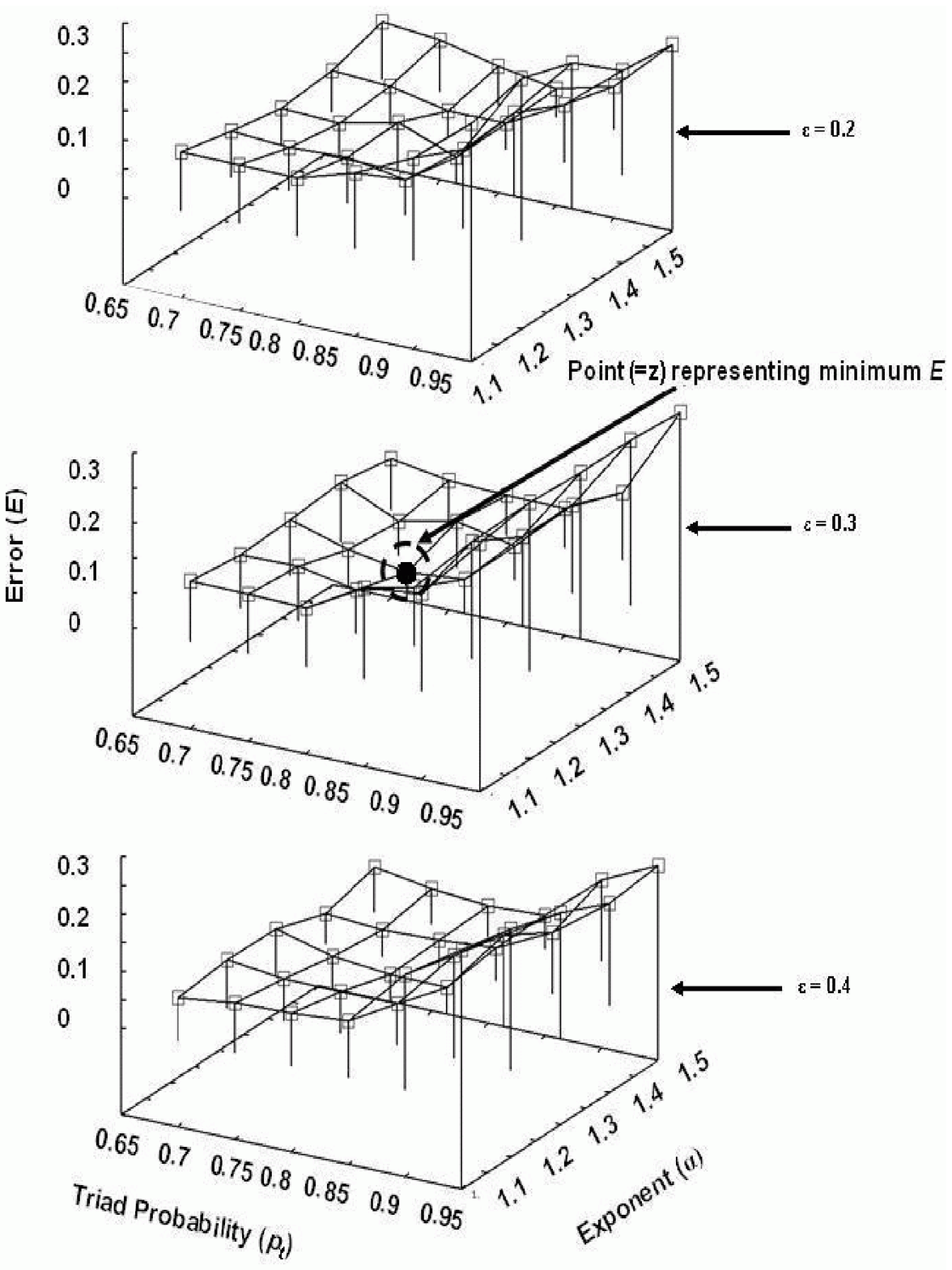}
\caption{The error $E$ for different values of the parameters $p_t$, $\alpha$ and $\epsilon$. The letter $z$ indicates 
the point where $E$ is minimum.}
\label{errors}
\end{center}
\end{figure}

We simulate the above synthesis model with different values of $\alpha$ ranging from 1.1 to 1.5 (in steps of 0.1) and 
$p_t$ ranging from 0.70 to 0.95 (in steps of 0.05) and compute the error $E$ in each case. Figure~\ref{errors} shows 
the three dimensional plot with $p_t$ in the x-axis, $\alpha$ in the y-axis and $E$ in the z-axis for three different 
values of $\epsilon$ which are 0.2, 0.3 and 0.4 respectively. We do not report the results for $\alpha~>~1.5$ since at 
larger values of $\alpha$ the error $E$ continuously keeps on rising mainly due to gelation (a condition where a 
single dominant ``gel" node is connected to all other nodes)~\cite{Ramasco:04}. Moreover, a reasonable amount of triad 
formation can only take place at higher values of $p_t$ and hence we do not report the results for $p_t~<~0.70$. 
Further, we do not choose larger values of $\epsilon$ since it is a smoothing parameter. The figure clearly shows that 
in our inspection range the minimum error ($E$~=~4.1$\%$) is achieved when the values of $p_t$, $\alpha$ and 
$\epsilon$ are 0.8, 1.3 and 0.3 respectively. We have also empirically observed that these parameters are stable in 
the sense that a slight perturbation in them brings only a negligible change in the results.     

\subsection{Simulation Results}

We plug in the best values (corresponding to minimum $E$) of the parameters $p_t$, $\alpha$ and $\epsilon$ as obtained 
in the earlier section in Algorithm~2 to synthesize PlaNet$_{syn}$ and subsequently PhoNet$_{syn}$. The results of 
this simulation are provided below.

\begin{figure}
\begin{center}
\includegraphics[width=3.3in]{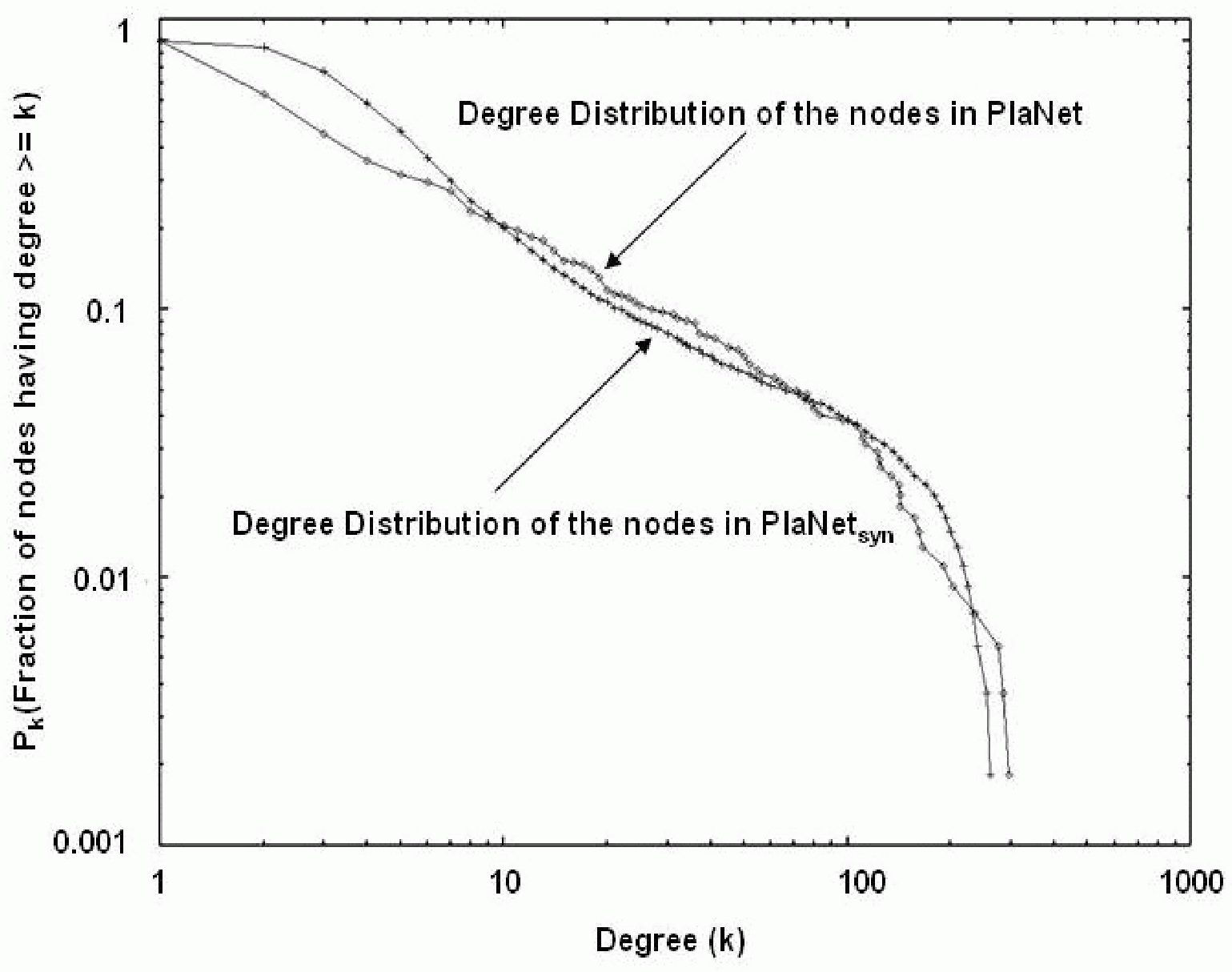}
\caption{Degree distribution of PlaNet$_{syn}$ along with that of PlaNet in log-log scale. The values of the 
parameters $p_t$, $\alpha$ and $\epsilon$ are 0.8, 1.3 and 0.3 respectively.}
\label{pltr}
\end{center}
\end{figure}

\begin{figure}
\begin{center}
\includegraphics[width=3.3in]{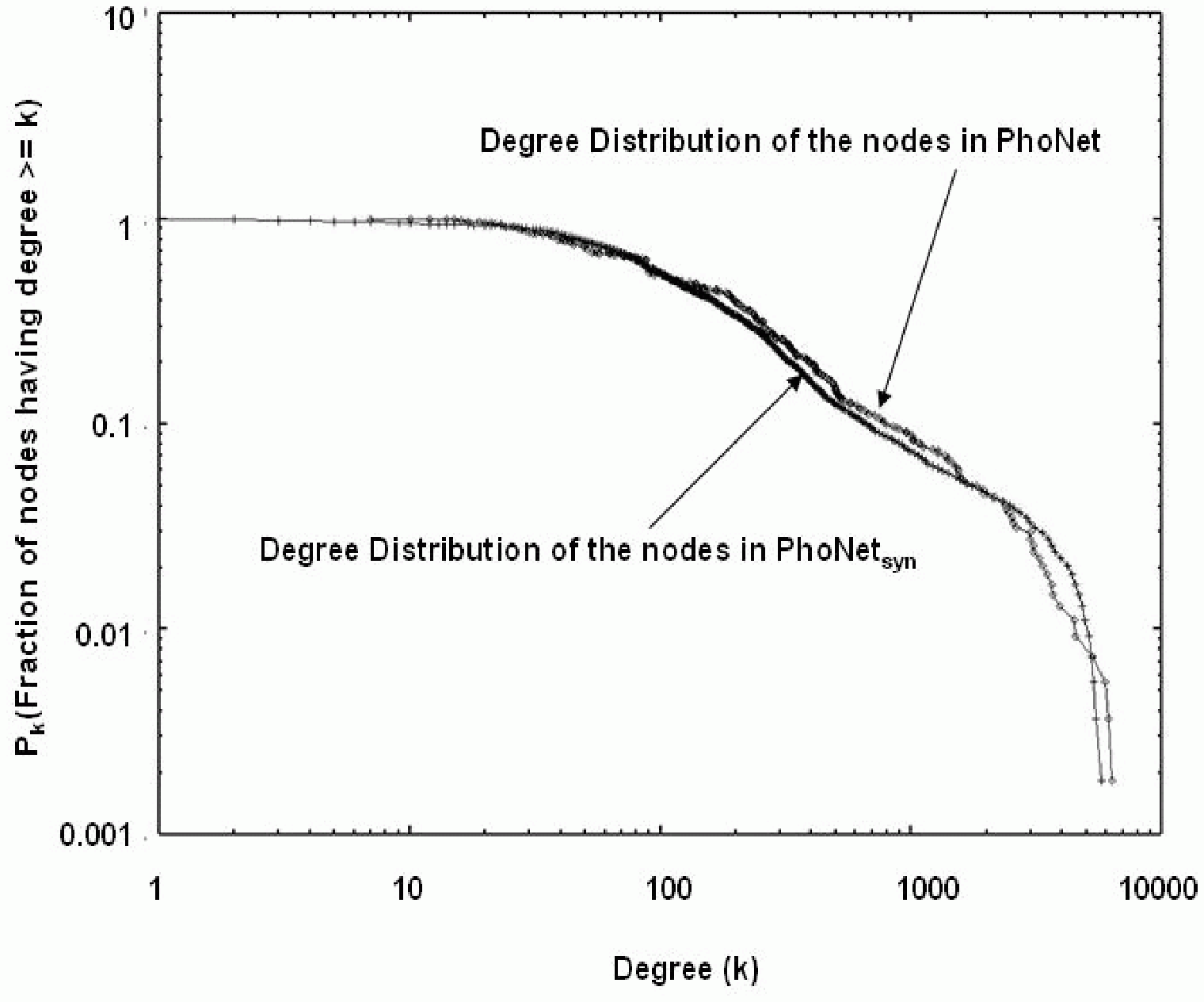}
\caption{Degree distribution of PhoNet$_{syn}$ along with that of PhoNet in log-log scale. The values of the 
parameters $p_t$, $\alpha$ and $\epsilon$ are 0.8, 1.3 and 0.3 respectively.}
\label{phtr}
\end{center}
\end{figure}

Figure~\ref{pltr} shows the degree distribution of the consonant nodes of PlaNet$_{syn}$ in comparison with that of 
PlaNet. The mean error between the two distributions is 0.04 approximately and is therefore a deterioration from the 
earlier result. This is mainly due to the trade-off involved in the definition of $E$. Nevertheless, the average 
clustering coefficient of PhoNet$_{syn}$ in this case is 0.85 as compared to 0.89 of PhoNet. Moreover, in this process 
the mean error between the degree distribution of PhoNet$_{syn}$ and PhoNet (as illustrated in Figure~\ref{phtr}) has 
got reduced dramatically from 0.45 to 0.03.

The above results equivocally indicate that for a good choice of the parameters (described earlier) the refined 
version of the synthesis model not only reproduces the degree distribution of PlaNet but also the degree distribution 
as well as the average clustering coefficient of PhoNet to a very close approximation.

\section{Conclusions, Discussion and Future Work}\label{conc}

In this article we have analyzed and synthesized the consonant inventories of the world's languages through a complex 
network approach. We dedicated the preceding sections for the following,
\begin{itemize}
\item Propose complex network representations of the consonant inventories, namely PlaNet and PhoNet,
\item Provide a systematic study of some of the important structural properties of PlaNet and PhoNet,
\item Develop a synthesis model for PlaNet based on preferential attachment where the consonant inventory size 
distribution is known {\em a priori},
\item Refine the synthesis model so that it not only  mimics the degree distribution of the consonant nodes of PlaNet 
as well as PhoNet but also reproduces the clustering coefficient of PhoNet to a very close approximation.
\end{itemize}  

Until now we have been mainly dealing with the computational aspects of the distribution of consonants over the 
languages rather than exploring the real world dynamics that gives rise to such a distribution. Language is a 
constantly changing phenomena and its present structure is determined by its past evolutionary history. The 
sociolinguist Jennifer Coates remarks that this linguistic change occurs in the context of linguistic heterogeneity. 
She explains that ``. . . linguistic change can be said to have taken place when a new linguistic form, used by some 
sub-group within a speech community, is adopted by other members of that community and accepted as the 
norm."~\cite{Coates:93}. In this process of language change (at the microscopic level), consonants belonging to 
languages that are more prevalent among the speakers in one generation have higher chances of being transmitted to the 
speakers of languages of the subsequent generations~\cite{Abrams:03,Blevins:04}. In the mesoscopic level this 
heterogeneity in the choice of the consonants manifests itself as preferential attachment. 
Further, if two consonants largely co-occur in the languages of one generation, it is highly likely that they will be 
transmitted together in the languages of the following generations~\cite{Blevins:04}. The aforementioned phenomenon is 
what is reflected through the formation of triads discussed in the earlier section. It is interesting to note that 
whereas triad formation among consonants takes place in a top-down fashion as a consequence of language change over 
linguistic generations, the same happens in a social network in a bottom-up fashion where actors come to know one 
another through other actors and thereby slowly shape the structure of the whole network. Moreover, unlike in a social 
network where a pair of actors can regulate (break or acquire) their relationship bonds, if the co-occurrence bond 
between two consonants breaks due to pressures of language change it can be never acquired again\footnote{There is a 
very little chance of reformation of the bond only if by coincidence the speakers learn a foreign language which has 
in its inventory one of the consonants lost (from the inventory of the native language of the speaker) in the process 
of language change.}. Such a bond breaks only if one or both of the consonants are completely lost in the process of 
language change and is never formed in future since the consonants that are lost do not reappear again. In this 
context, Darwin in his book, {\em The descent of man}~\cite{Darwin:71}, writes ``A language, like a species, when once 
extinct never reappears."

Although the directions of growth in a social network is different from the networks discussed in this article, both 
of them target to achieve the same configuration. It is mainly due to this reason that the principle of preferential 
attachment along with that of triad formation is able to capture the self-organizing behavior of the consonant 
inventories.     

In this article we have mainly dealt with the unlabeled synthesis (since the consonant nodes are unlabeled in our 
synthesis model) of the occurrence and co-occurrence networks of consonants. However, the work can be further extended 
in the directions of a labeled synthesis of the consonant networks (and hence consonant inventories). We look forward 
to accomplish the same as a part of our future work.

\end{document}